# Structural evolution of iodine on approach to the monatomic state

Elena Bykova[1,3], Iskander G. Batyrev[2], Maxim Bykov[1,4], Eric Edmund[1], Stella Chariton[5], Vitali B. Prakapenka[5], and Alexander F. Goncharov[1,4]

[1] Earth and Planets Laboratory, Carnegie Institution for Science, Washington, DC 20015, USA

[2] U.S. Army Research Laboratory, RDRLWML-B, Aberdeen Proving Ground, Maryland 21005, United States

[3] Bayerisches Geoinstitut, University of Bayreuth, Universitätsstrasse 30, D-95447, Bayreuth, Germany

[4] Institute of Inorganic Chemistry, University of Cologne, Greinstrasse 6, 50939 Cologne, Germany

[5] Center for Advanced Radiation Sources, The University of Chicago, Chicago, Illinois 60637, USA

**We applied single-crystal X-ray diffraction and Raman spectroscopy in a diamond anvil cell up to 36 GPa and first principles theoretical calculations to study the molecular dissociation of solid iodine at high pressure. Unlike previously reported, we find that the familiar *Cmce* molecular phase transforms to a *Cmc*2$_1$ molecular structure at 16 GPa, and then to an incommensurate dynamically disordered *Fmmm*(00γ)*s*00 structure at 20 GPa, which can be viewed as a stepwise formation of polymeric zigzag chains of three iodine atoms following by the formation of the dynamically dissociated, incommensurately modulated *i-Fmmm* phase, and the truly monatomic *Immm* phase at higher pressures.**

Understanding pressure driven dissociation of simple diatomic molecules is important for a range of topics including materials behavior under extremes and composition of planetary interiors. Such transformations, which are commonly preceded or accompanied by metallization, occur at very high pressures in the crystals made of light molecules (e.g., $H_2$ and $F_2$), where such investigations are very challenging (1, 2). However, molecular dissociation and band gap closure occurs at much lower pressures for heavier molecules. In fact, these phenomena in halogens $Cl_2$, $Br_2$, and $I_2$ have been thoroughly investigated revealing a common behavior (3-5), where molecular dissociation is reported to occur in steps associated with a series of phase transitions.

Iodine is investigated most extensively largely because it transforms to a metallic state at 14-24 GPa and experiences molecular dissociation at above 21 GPa (6-11)—conditions, which are easily accessible with the diamond anvil cell technique. At low pressures, iodine crystallizes in an orthorhombic *Cmce* structure (3) (phase I), which consists of flat layers, formed as associations of zigzag molecular chains (Fig. 1). Above 26 GPa, this structure transforms into a monatomic *Immm* lattice (phase II), where iodine atoms occupy the corner sites forming a metallic body-centered orthorhombic (BCO) single-atom unit cell (8). However, careful investigations (12) showed the existence of an intermediate phase (V) at 24-26 GPa, which can coexist with the *Immm* structure at 26-30 GPa. This intermediate structure can be viewed as a face-centered orthorhombic lattice based on the main diffraction peaks, but the occurrence of satellite reflections from both sides of the main peaks indicates the presence of an incommensurate modulation wave along the *a*-axis, which shifts the atoms along the *b*-axis. In this structure, there are three- or four- atom chains in the *ab* plane, with a continuous distribution of the nearest-neighbor interatomic distances covering a range of approximately 0.26 Å. Raman spectroscopy demonstrated the presence of a specific low-frequency soft mode for the modulated structures, called the amplitude mode (AM), which corresponds to transverse atomic vibrations and preserves the symmetry of the modulation wave (5). No Raman signal has been detected in phase II as expected from the Raman selection rules.

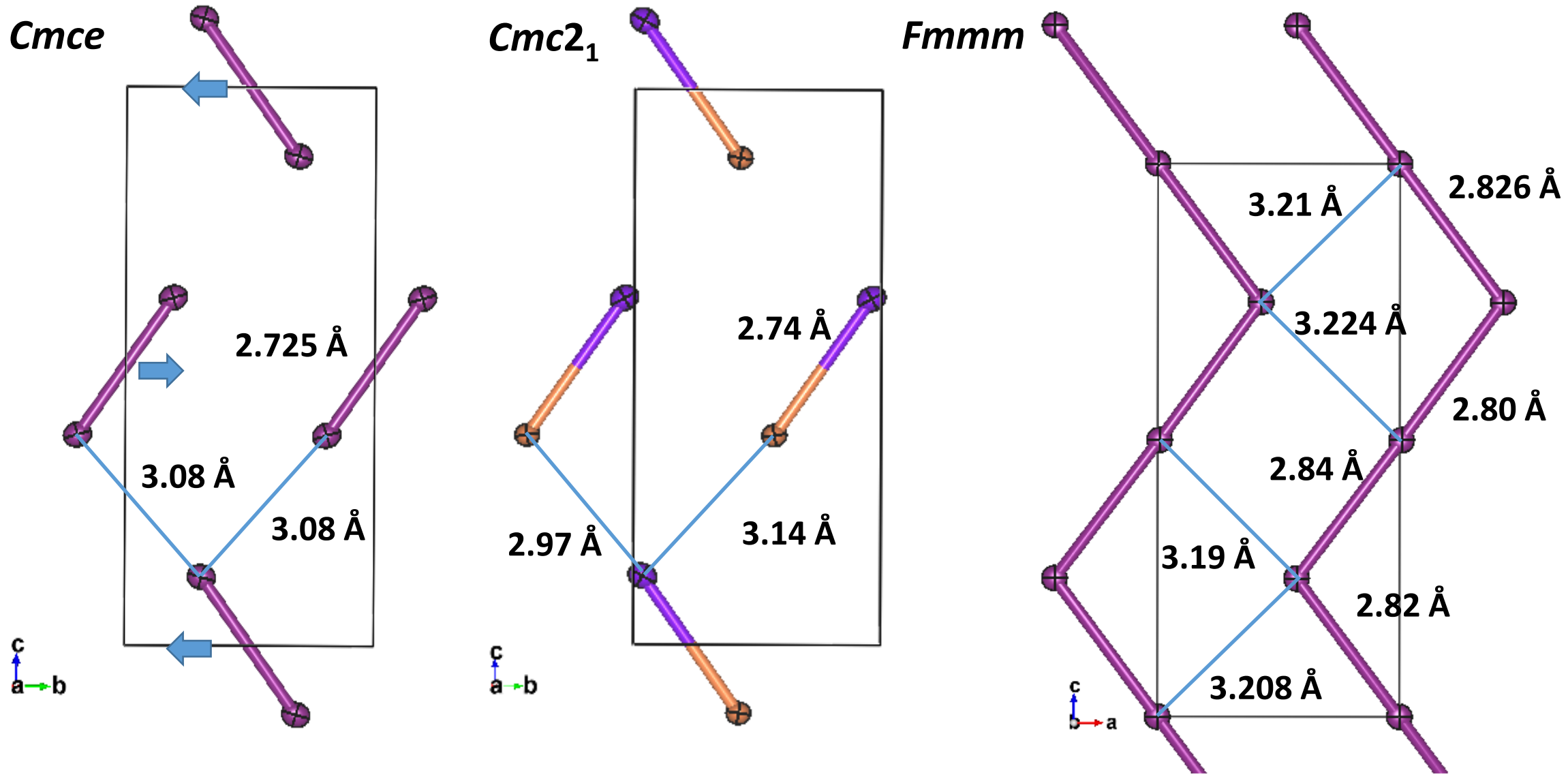

**Fig. 1.** Fragments of crystal structures of *Cmce* (I), *Cmc*2₁ (VI), and *Fmmm*(00γ)*s*00 (VII) phases determined in this work (Tables S1 and S2, of Supplemental Material (13)) projected to the *bc*(*ac* for *Fmmm*) plane. The atoms occupying different crystallographic sites are shown in different colors. The atoms are represented by the displacement ellipsoids (at 75% probability) as they occur after the structural refinement. Arrows show the directions of shifts of the molecular layers that occur at the *Cmce* to *Cmc*2₁ transition. *Fmmm*(00γ)*s*00 structure presentation is approximate (small unit cell) (see detailed structural results in Table S2, Supplemental Material (13)).

In the stepwise molecular dissociation scenario presented above, it remains puzzling if molecular phase I metallizes via band overlap and whether there is any indication of its instability toward the transformation into modulated phase V. Indeed, theoretical calculations predict such instability within *Cmce* structure due to a coupling of the bond charge density with a transverse optical phonon (14). Raman measurements of phase I (5, 15) demonstrated the presence of $2A_g$+$2B_{3g}$ in-plane stretching (S) and librational (L) modes, out of which $A_g$(L) mode turns over and becomes a soft mode above 13 GPa. In the same pressure range, additional Raman modes appear that cannot be explained within the Raman selection rules of *Cmce* structure. In addition, Mössbauer experiments demonstrated an anomaly at 16 GPa suggesting a phase transformation (16), while X-ray absorption spectroscopy showed a hint of an increase in the interamolecular bond distance in the same pressure range (4). Theoretical first-principles calculations (17) suggested that there is another molecular phase (phase I′), which has *C*2/*m* symmetry; it was proposed to coexist with *Cmce* phase I (17). This phase has two types of $I_2$ molecules in the unit cell with different intramolecular lengths offering an explanation to the extra Raman bands, which appear above 14 GPa.

Here we present the results of concomitant synchrotron single-crystal (SC) X-ray diffraction (XRD) and Raman measurements, which provide a detailed structural description of the multi-stage molecular dissociation of iodine. Our experiments unequivocally identify intermediate phases - a *Cmc*$2_1$ molecular phase (hereafter $I_2$-VI), and an incommensurate *Fmmm*(00γ)*s*00 phase (hereafter $I_2$-VII) characterized by the loss of well-defined stable molecules, which provide key missing links between stable molecular $I_2$-I and dynamically dissociated $I_2$-V. Our first-principles theoretical calculations show that *Cmc*$2_1$ $I_2$-VI phase is energetically competitive, dynamically stable and confirm the presence of the additional Raman bands as due the reduction in symmetry upon transition to the *Cmc*$2_1$ phase.

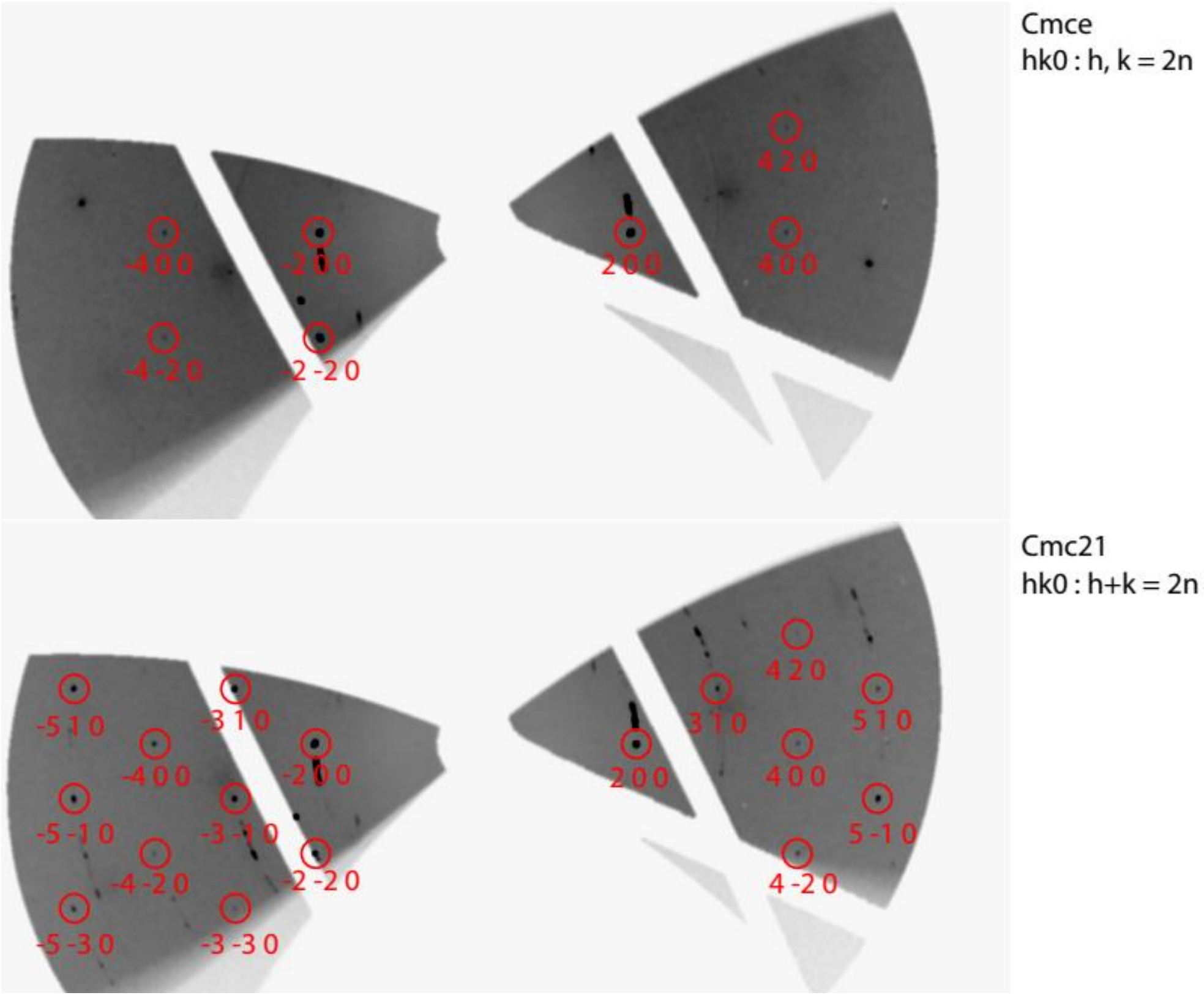

**Fig. 2.** Reconstructed reciprocal lattice planes of iodine at 15 GPa (upper panel) and at 17 GPa (lower panel). The observed diffraction spots from the sample have been indexed and used to determine the structure of *Cmce* ($I_2$-I) and *Cmc*$2_1$ ($I_2$-VI) phases, respectively (see details in Table S1 of Supplementary Materials (13)). The X-ray extinctions rules for these phases are depicted.

Up to 15 GPa SC XRD patterns confirm the previously reported *Cmce* structure (Table S1 of Supplementary Information (13)). However, SC XRD measurements at 17 GPa demonstrate a difference in the X-ray extinction rules. Reflections (h k 0), with h=2n+1 and k = 2n+1 (e.g. (3 1 0), (5 1 0) and others) appear in the diffraction patterns (Fig. 2). These weaker reflections cannot be observed in powder XRD patterns as they are much weaker than the main ones (<0.1%) and interfere with stronger reflections, so this symmetry change was previously overlooked. A new structure can be indexed with the $Cmc2_1$ space group, which represents a subgroup of the *Cmce* group. Hereafter, we call this phase as $I_2$-VI (cf. Ref. (18)). The structural distortion in the $Cmc2_1$ structure is very subtle (Fig. 1). It can be understood as a small shift of layers of collinear molecules along the *b*-axis, which diminishes the lattice symmetry.

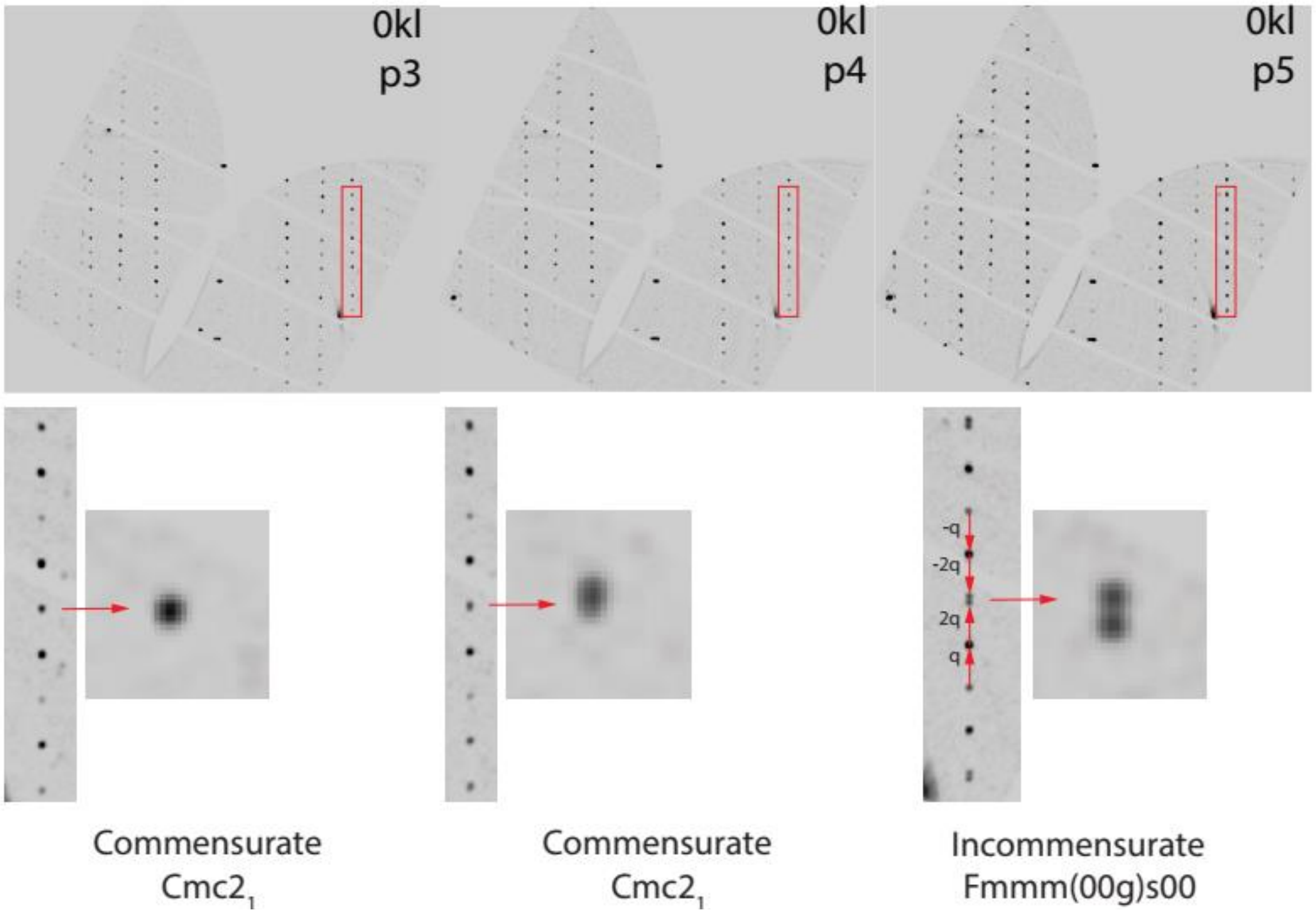

**Fig. 3.** Reconstructed reciprocal lattice planes of iodine at 17, 19, and 20.8 GPa (from left to right). Insets below show enlarged views of Bragg peaks marked by red rectangles. Splitting of the zoomed in peak manifests the occurrence of incommensurate phase at 20.8 GPa.

On further compression, above 20 GPa, yet another subtle change in symmetry occurs (Figs. 1, 3). At this point, a stable structural solution within $Cmc2_1$ space group is not possible. A sequence of diffraction spots from (0kl) lattice planes (Fig. 3) show a splitting of several reflections along the c* direction at 20.8 GPa, manifesting the occurrence of an incommensurate lattice with a modulation vector of q=0.4837(2). An incommensurate phase has *Fmmm*(00γ)s00 structure, which is isosymmetrical with phase V at higher pressures. The existence of this phase is consistent with Ref. (18), where this phase $I_2$ was reported  at 16-23 GPa. In contrast, we find no incommensurate structure below 20 GPa, where commensurate *Cmc*2$_1$ $I_2$-VI is stable. Both commensurate *Cmc*2$_1$ and *Cmce* are the $a \times b \times 2c$ supercell subgroups of *Fmmm*(00γ)s00 with $\gamma = 0.5$ and various initial phases of modulation $t_0$ (1/8 for *Cmce* and general for *Cmc*2$_1$).

In incommensurate $I_2$-VII phase, the nearest-neighbor interatomic distances are modulated (Fig. S1 of Supplementary Information (13)) spreading a wide range, which covers typical intra-to-intermolecular I-I distances. This phase consists of a dynamic mixture of molecular $I_2$ and polymeric zigzag chains of three I atoms (see also Ref. (18)) in the *ac*-plane (Fig. S2 of Supplementary Information (13)).

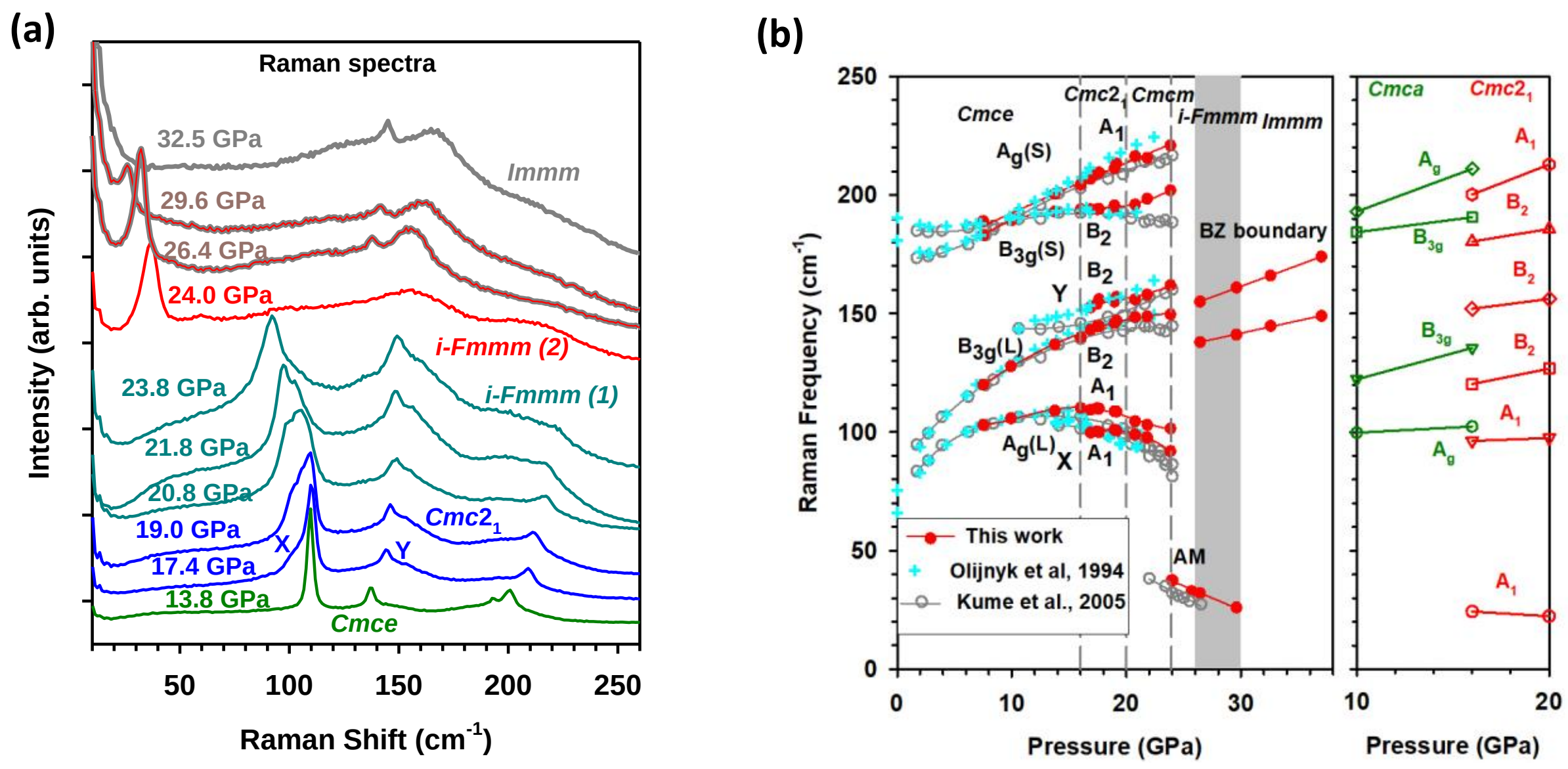

**Fig. 4.** Raman spectroscopy data. Panel (a) shows a sequence of Raman spectra across the phase transitions on pressure increase. The phase symmetries are labeled, and the traces are color coded. Modes labeled as X and Y appear in *Cmc*2$_1$ phase and gain intensity. Panel (b) shows the measured here pressure dependencies of the Raman frequencies and compared to previous investigations (5, 15) (left panel) as well as theoretically calculated (right panel) in this work.

Our concomitant Raman measurements below 24 GPa demonstrate smooth changes with pressure (Fig. 4), which are generally consistent with previous observations (5, 15). Above 16 GPa, where *Cmc*2$_1$ $I_2$-VI is documented in XRD, new peaks appear, dubbed X and Y in the literature, with the frequencies below $A_g$(L) and above $B_{3g}$(L) modes, respectively. The appearance of these modes can be understood by the Raman selection rule modification at the *Cmce*→*Cmc*2$_1$ phase transition (Table S4 of Supplementary Information (13)); the most prominent difference in *Cmc*2$_1$ phase VI is the relaxation of the rule of mutual exclusion for the Raman and IR active modes. According to group-theory predictions, all optical modes are Raman-active in phase VI, out of which the most intense $3A_1$+$3B_2$ in-plane modes are expected to be observed in Raman spectra compared to $2A_g$+$2B_{3g}$ modes of *Cmce* phase I. We note that X and Y modes are coupled to $A_g$(L) and $B_{3g}$(L) originated modes, respectively. This is evident from their frequency curves vs. pressure (Fig. 4(b)), which show characteristic avoided crossing dependencies and an exchange in intensities correlated with the frequency approaching as in the case of Fermi resonance. This behavior shows that the coupled modes are of the same symmetry, namely $A_1$ and $B_1$ modes of *Cmc*$2_1$ phase (Fig. 4(b)). Thus, the appearance and behavior of the X and Y modes are well understood by the *Cmce*→*Cmc*$2_1$ phase transformation; these modes originate from IR active $B_{1u}$+$B_{2u}$ modes of *Cmce* phase, which become Raman active in *Cmc*$2_1$ phase (Table S3 of Supplementary Information (13)). These modes correspond to translational motions of $I_2$ molecules along the *b* and *c* axes, which couple with librational motions in phase VI.

Only subtle changes (mainly peak broadening) occur in the Raman spectra above 20 GPa, where our XRD data indicate a transition into an incommensurate *Fmmm*(00γ)s00 $I_2$-VII phase. The presence of vibrational modes characteristic for molecular phases is likely because this phase still consists of short lived $I_2$ molecules demonstrating fluxional behavior, which is supported by MD simulations of Ref. (18). This behavior is consistent with the incommensurate nature of this phase, where molecular breakdown and recombination are associated with the modulation wave propagating through the crystal. It is the instantaneous existence of molecules that leads to a Raman response similar to that of *Cmce* and *Cmc*$2_1$ phases. At 23.8 GPa, the $A_1$ mode originated from the X-mode becomes a dominant mode in the spectrum; this mode is a soft one corresponding to librational motions of $I_2$ molecules with the eigenvector, which captures the pathway to a monatomic lattice. The observed here mode behavior is qualitatively consistent with the first-principles theoretical calculations (Fig. 4 and Fig. S3 of Supplementary Information (13)).

Our first-principles theoretical calculations show that *Cmc*$2_1$ phase is energetically competitive with respect to *Cmce* and *C2/m* in the pressure range above 15 GPa , where *Cmce* phase is reported to transform in this work and Refs. (17, 18) (Fig. S4 of Supplementary Information (13)). However, within GGA-PBE approximation, *Cmc*$2_1$ phase is substantially more stable than the experimental *Cmce* ground-state phase below 15 GPa (see also Ref. (19)), which clearly contradicts the experiments. This apparent inconsistency is removed if M06-L functional is used as the enthalpies of *Cmce* and *Cmc*$2_1$ phases become nearly degenerate in the whole pressure range of interest. All these phases are dynamically stable at 5-30 GPa as witnessed by calculations of the phonon dispersion curves (Fig. S5 of Supplementary Information (13)). GGA-PBE calculations of the electronic band structure show that the molecular phases determined here are semiconducting with narrow band gaps decreasing with pressure, which are expected to close at 26 GPa (Fig. S6 of Supplementary Information (13)).

At 24 GPa, XRD patterns and Raman spectra change abruptly. SC XRD performed in this work find a modulated structure (Figs. S7, S8 and Table S3 of Supplementary Information (13)) as reported in Ref. (12) based on powder diffraction data (phase V). Unlike Ref. (12), we find the structure of phase V to be *Fmmm*(00γ)*s*00 (cf. *Fmm*2(*a*00)0*s*0 in Ref. (12)) in agreement with the more recent analysis (20). We also find that the modulation vector value decreases with pressure (Fig. S9) in agreement with Refs. (12, 20). However, we additionally find, based on SC XRD measurements on high-quality laser annealed crystals (Fig. S7 of Supplementary Information (13)), that there is a tiny atomic displacement along the *c*-axis (Table S3 of Supplementary

Information (13)). Overall, these results definitively establish the structure of a modulated phase V proposed in Ref. (12). However, our data do not support the tetragonal 5D modulated structure (*I*4/*mmm*(αα0)000*s*(−αα0)0000) proposed in Ref. (18). As can be seen on the reconstructed precession image (Fig. S7 of Supplementary Information (13)), there is only one modulation vector, which can describe all the satellite reflections.

Raman spectra above 24 GPa do not show any modes of the lower pressure molecular phases I, VI, and VII. Instead, a strong soft Raman mode appears at low frequencies, which has been previously identified as an AM (5) (Fig. 4(b)). These observations confirm that the modulated phase V is incommensurate, where the Brillouin zone center vibrational modes of the parent phase are forbidden. Above 26 GPa, two new Raman excitations, a narrow and a broad at nearly 150 $cm^{-1}$ appear and shift to higher frequencies with pressure (Fig. 4(b)). Monatomic (BCO) phase II coexists with phase V in this regime, identified in powder XRD measurements. Above 30 GPa, the AM mode disappears, and XRD measurements show a single-phase BCO structure. Based on our theoretical calculations (Fig. S10), the Raman bands observed in this phase correspond well in frequency and pressure shift to the Brillouin zone boundary transverse acoustic modes near the R, W, and T points, where the corresponding dispersion curves are flat yielding sharp maxima in the phonon density of states at 4 THz (134 $cm^{-1}$). The mechanism of their Raman activity (e.g., defect induced) is not clear at this stage.

Our results show the presence of an intermediate molecular phase VI and fluxional mixed molecular-zigzag incommensurate phase VII between a low-pressure molecular phase I and a modulated and dynamically dissociated phase V. Phase VI has $Cmc2_1$ orthorhombic symmetry, which can be described as a small distortion of *Cmce* low-pressure phase. This new phase can be viewed as a slightly dissociated (with longer intramolecular distance (Figs. 1, 5) molecular phase, where intermolecular coupling has been modified to approach zigzag chains of three iodine atoms. Previous experiments and theoretical calculations proposed a monoclinic *C*2/*m* structure in this regime (17), but our XRD experiments clearly rule out this phase. In a recent work, it has been proposed that phase VI is a modulated incommensurate phase (18); however, our SC XRD data clearly show that no satellite reflections are observed below 20 GPa. In a qualitative agreement with the results of previous works (12, 18, 20), we find the structure of phase V at 24-30 GPa to be modulated and incommensurate (Figs. S7-S9 of Supplementary Information (13)); however, the symmetry and the dimensionality are different of that proposed in Ref. (18).

Our new phase identification is consistent with Mössbauer experiments (16), which proposed a change in symmetry based on discontinuous changes in the electric field gradient (EFG) parameters. We theoretically computed these parameters in *Cmce* and $Cmc2_1$ phases and found discontinuous changes in EFG, which are qualitatively consistent with the Mössbauer observations (Fig. S11 of Supplementary Information (13)).

The *Fmmm*(00γ)s00 incommensurate mixed molecular-zigzag phase VII differs from incommensurate phase V in that the latter one can be viewed as totally dissociated, consisting of $I_4$ and $I_5$ zigzag atomic chains (Fig. S1), while phase VII still has a large disparity between intra and intermolecular bond lengths, which results in formation of zigzag $I_3$ chains (cf. Ref. (18)), while some $I_2$ molecules still remain. The modulation vector of phase VII is substantially larger than that of phase V (Fig. S9). The interatomic distances in both modulated phases V and VII greatly vary (Fig. 5, Fig. S1), while the averaged distance goes down with pressure due to the volume compression. The determined here phase sequence Commensurate (*Cmce*)→ Commensurate (*Cmc*$2_1$)→ Incommensurate *Fmmm*(00γ)s00 (γ ~ 0.5) → Incommensurate *Fmmm*(00γ)s00 (γ ~ 0.25) → *Immm* shows a seamless pathway for molecular dissociation of $I_2$.

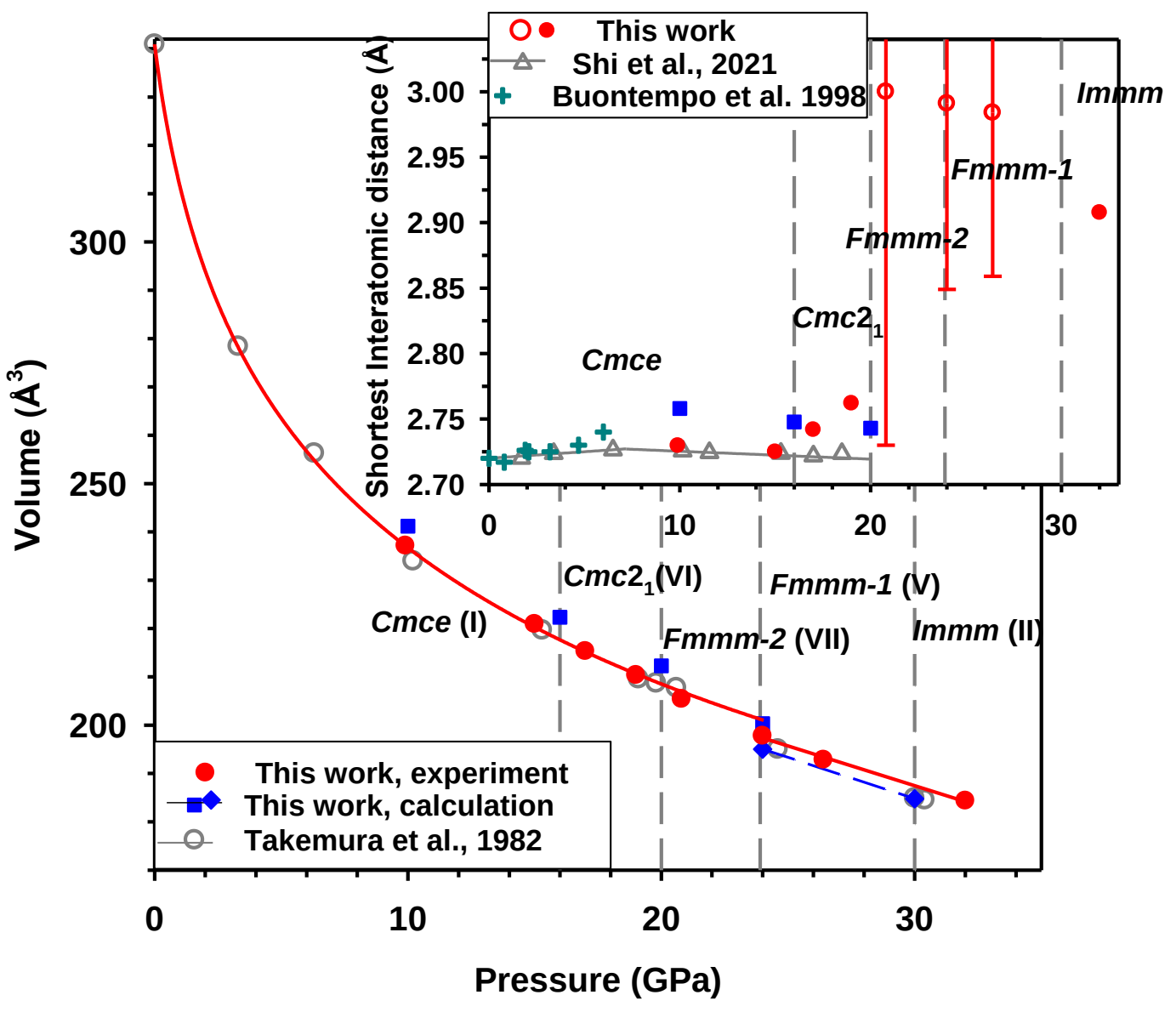

**Fig. 5.** Volume compression of iodine. The inset shows the nearest interatomic distance vs pressure data. The results of this work (filled symbols, the uncertainty is smaller than the symbol size) are compared to previously reported data (open symbols and crosses) from Refs. (4, 8, 21). The solid line extrapolated to the transition into the modulated *Fmmm* phase is a Vinet fit to our data ($V_0$=340.9(2) Å$^3$, $K_0$=8.1(5), and $K_0'$=7.0(3), where $K_0$ and $K_0'$ are the bulk modulus and its pressure derivative at P=0). The vertical bars in inset show the spread of the nearest I-I distances in incommensurate phases.

Our XRD data also address the volume change at metallization and molecular dissociation transition (Fig. 5). There is no measurable volume change at the *Cmce* → *Cmc*$2_1$ (VI)→ *Fmmm*(00γ)s00 (VII) phase transitions. However, there is a volume collapse of 1.8% that occurs at the isosymmetrical VII-V transition (both have *Fmmm*(00γ)s00 symmetry) (cf. Ref. (12)). This is smaller than that theoretically computed here using DFT calculations using GGA-PBE functional (4.2%). It has been previously suggested that metallization occurs in a molecular state thus preceding dissociation (4). Based on our XRD results and theoretical calculations of the electronic structure (Figs. S3, S4), we speculate that the *Cmc*$2_1$ → *Fmmm*(00γ)s00 (VII) phase transitions at 20 GPa can be directly connected to metallization reported at 14-24 GPa (9). In this case, there is no volume collapse associated with metallization. Molecular dissociation occurs stepwise from truly molecular *Cmce* and *Cmc*$2_1$ to dynamically disordered *Fmmm*(00γ)s00 and dynamically dissociated *i-Fmmm* phase. It would be of great interest to investigate if these results are applicable to $Br_2$ and $Cl_2$, which were reported to have a similar structural behavior.

Parts of this research were carried out at the GeoSoilEnviroCARS (The University of Chicago, Sector 13), Advanced Photon Source (Argonne National Laboratory). GeoSoilEnviroCARS is supported by the National Science Foundation—Earth Sciences (No. EAR-1634415). Use of the GSECARS Raman System was supported by the NSF MRI Proposal (No. EAR-1531583). The Advanced Photon Source is a U.S. Department of Energy (DOE) Office of Science User Facility operated for the DOE Office of Science by Argonne National Laboratory under Contract No. DE-AC02-06CH11357. We acknowledge support by the Army Research Office accomplished under the Cooperative Agreement No. W911NF-19-2-0172 and the Carnegie Institution of Washington. M.B. acknowledges the support of Deutsche Forschungsgemeinschaft (DFG Emmy-Noether project BY112/2-1). E.B. acknowledges financial support from the program 'Promotion of Equal Opportunities for Women in Research and Teaching' funded by the Free State of Bavaria.

## Supplementary Information

# Structural evolution of iodine on approach to the monatomic state

Elena Bykova[1,3], Iskander G. Batyrev[2], Maxim Bykov[1,4], Eric Edmund[1], Stella Chariton[5], Vitali B. Prakapenka[5], and Alexander F. Goncharov[1,4]

[1] Earth and Planets Laboratory, Carnegie Institution for Science, Washington, DC 20015, USA

[2] U.S. Army Research Laboratory, RDRLWML-B, Aberdeen Proving Ground, Maryland 21005, United States

[3] Bayerisches Geoinstitut, University of Bayreuth, Universitätsstrasse 30, D-95447, Bayreuth, Germany

[4] Institute of Inorganic Chemistry, University of Cologne, Greinstrasse 6, 50939 Cologne, Germany

[5] Center for Advanced Radiation Sources, The University of Chicago, Chicago, Illinois 60637, USA

*This pdf file contains Materials and Methods, Supplemental Figures S1-S11, Tables S1-S4, and Bibliography with References [22-27].*

*Corresponding author:*
*agoncharov@carnegiescience.edu*

## Materials and methods

### Experiments

The experimental procedure included concomitant SC XRD and Raman spectroscopy measurements (see details in the Supplementary Information) at 10-36 GPa in the Sector 13 (GSECARS) of the Advanced Photon Source, Argonne National Laboratory. Samples in the form of flakes were placed in a cavity of a diamond anvil cell formed in a central hole made in a preindented rhenium gasket, sealed to avoid sublimation, and then loaded by compressing Ne gas to 160 MPa at room temperature. Ruby served as an optical pressure gauge, while the equation of state of Ne [22] has been used to cross check pressure during XRD measurements. Compressed Ne serving as a pressure medium provided quasi-hydrostatic conditions in the high-pressure cavity, as witnessed by SC XRD data, which were of sufficiently high quality up to at least 27 GPa. At this pressure we performed laser annealing of the sample up to 1800 K to improve the quality of SC XRD data as will be presented below. Additional Raman measurements have been collected at Earth and Planets Laboratory of the Carnegie Institution for Science, before and after synchrotron XRD experiments.

### Theoretical calculations

First-principles theoretical calculations have been performed in *Cmce*, $Cmc2_1$, *C*2/*m*, *Immm* phases at selected pressures between 0 and 30 GPa, where these structures were optimized using norm-conserving pseudopotentials, GGA-PBE functional, and Grimme2 dispersion corrections[23]. Monkhorst-Pack grid size for k-points sampling of the Brillouin Zone (BZ) is 5x6x5 for all structures except $I_2$-II (*Immm*) at 30 GPa, where 6x7x6 sampling was used [24]. The phonon dispersion and phonon frequency calculations were performed using a finite displacement method implemented in the CASTEP code [25]. The Raman spectra were calculated using the formalism presented in Ref. [26]. Since application of GGA-PBE functional incorrectly predicts that *Cmce* phase is thermodynamically and dynamically unstable at 0 GPa [19], we performed calculations using Minnesota 2006 local functional [27] (M06-L) in *Cmce* and $Cmc2_1$ phases up to 30 GPa.

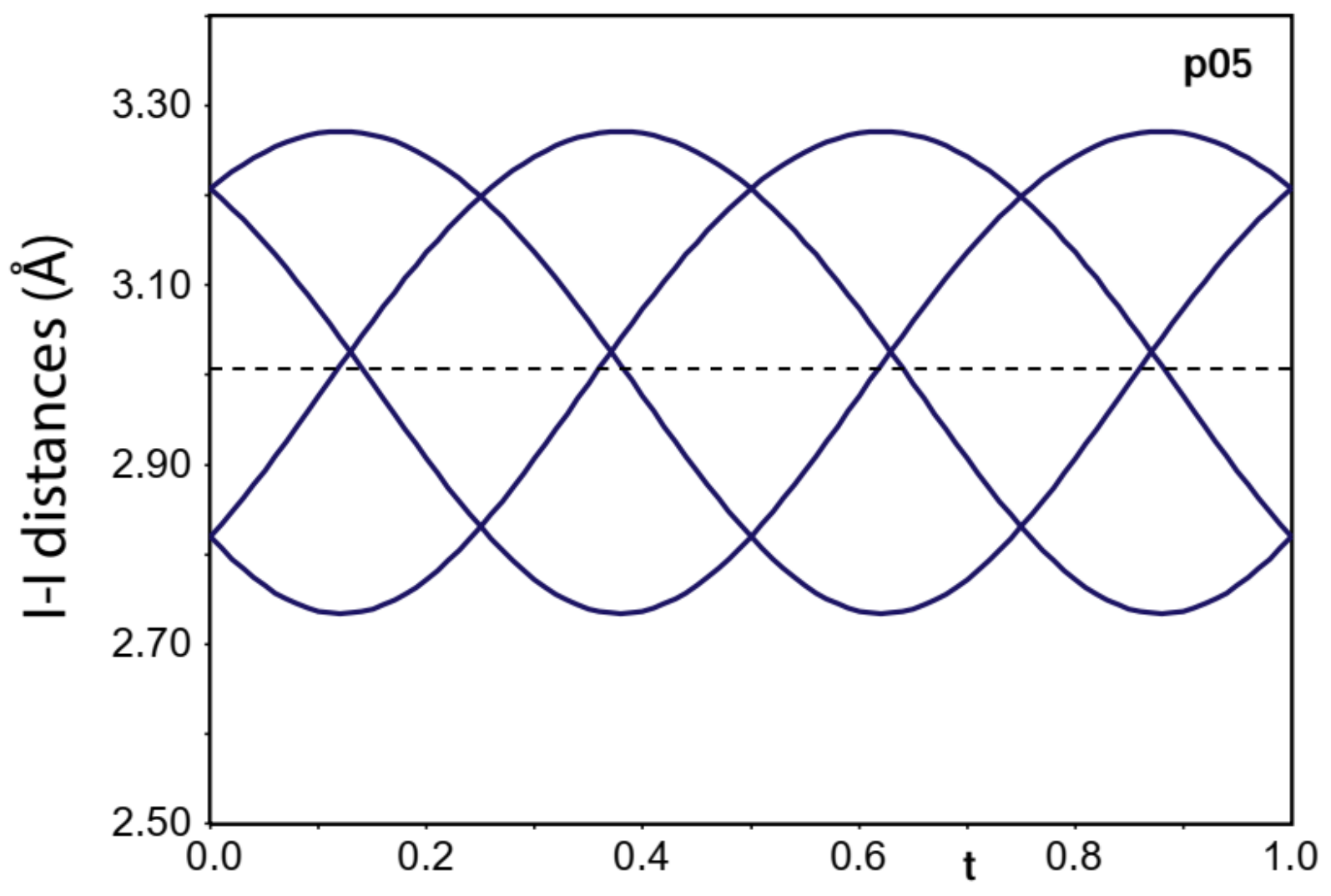

**Fig. S1.** The I-I nearest-neighbor interatomic distances as a function of the modulation vector of *Fmmm*(00γ)*s*00 $I_2$-VII phase determined in this work.

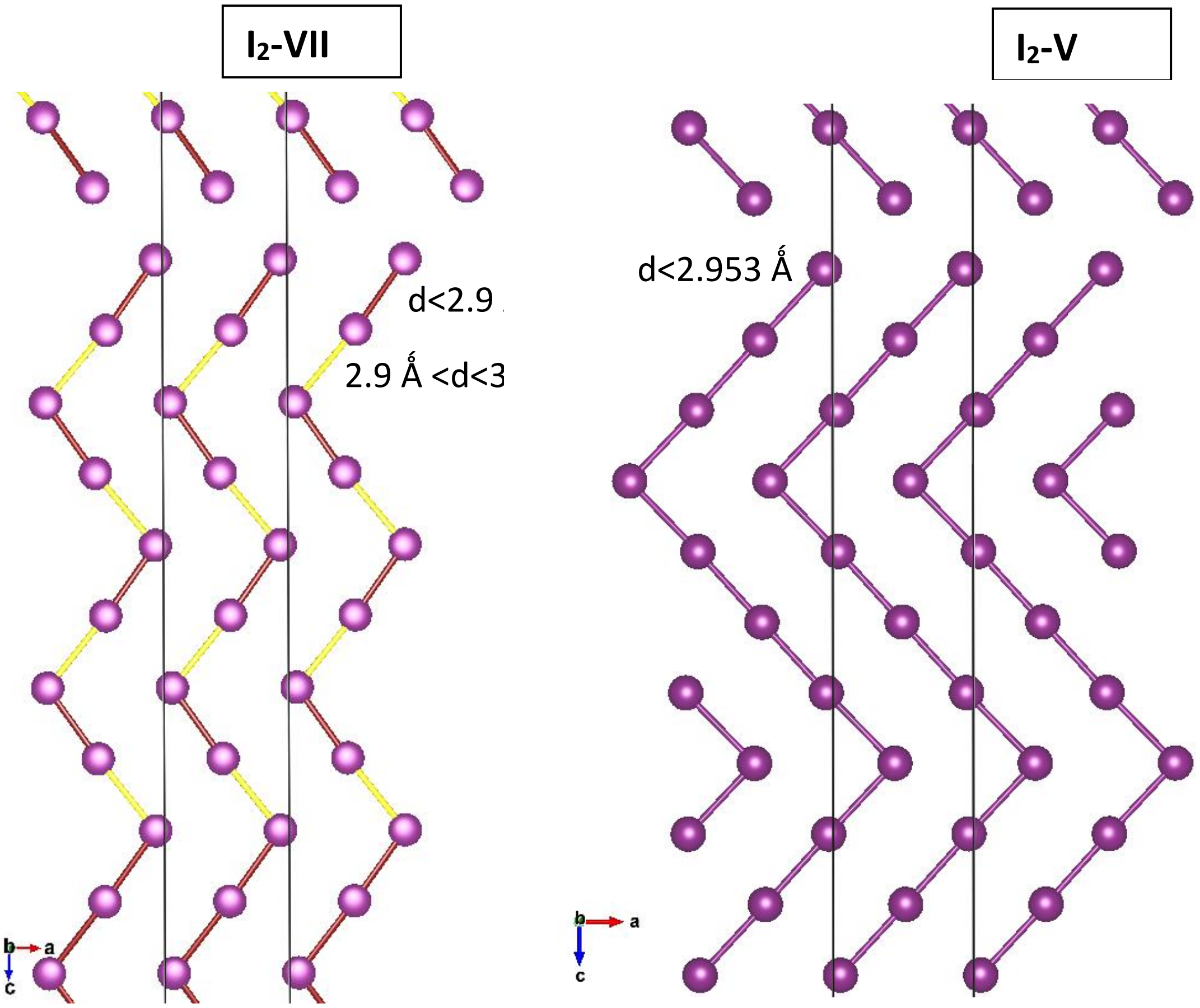

**Fig. S2.** Fragments of crystal structures of *Fmmm*(00γ)*s*00 $I_2$-VII and $I_2$-V phases determined in this work (Tables S2 and S3, respectively) projected to the *ac* plane.

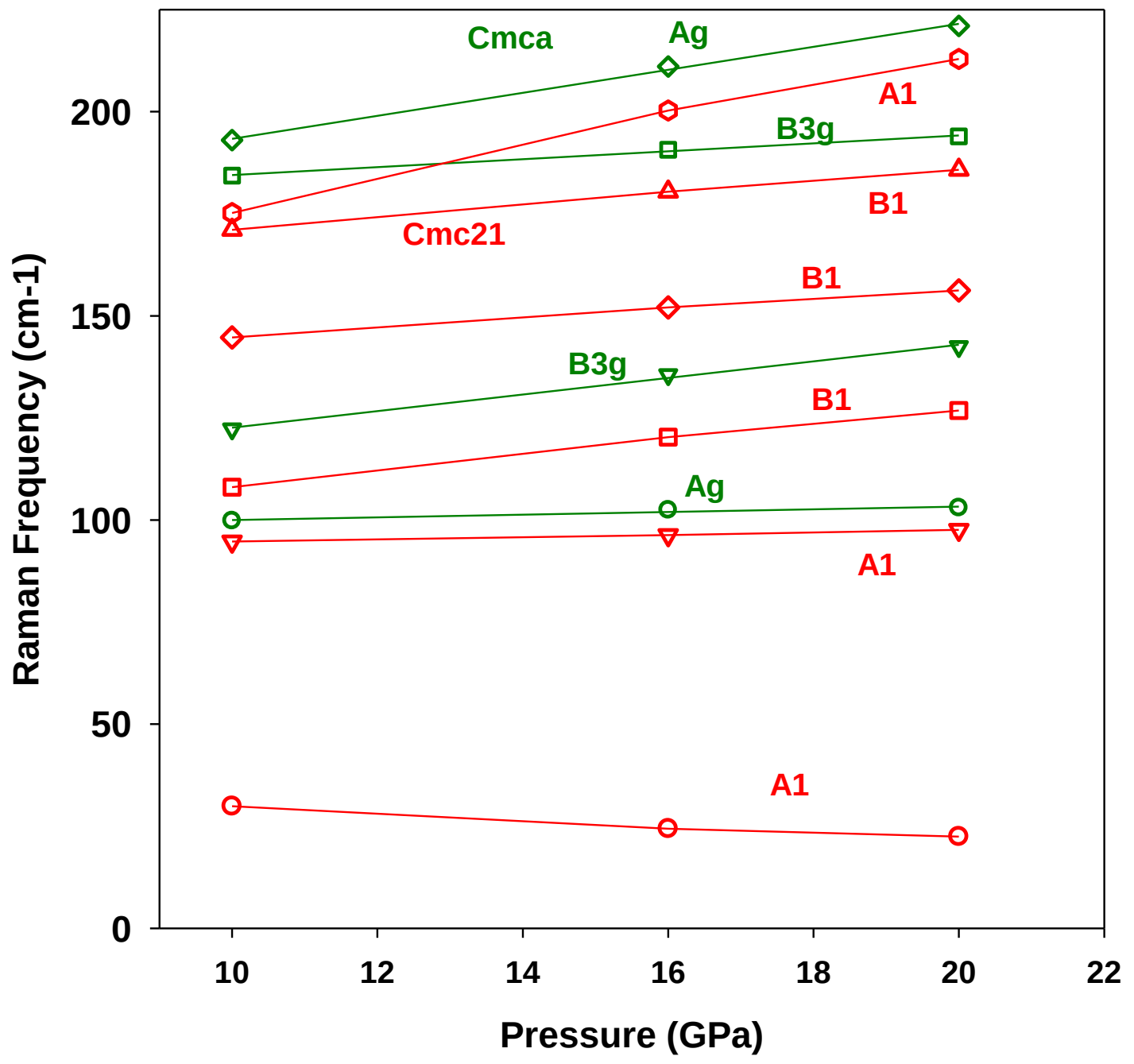

**Fig. S3.** Theoretically computed Raman frequencies vs pressure determined here from the first principles. The irreducible representations are changed to be consistent with the axes name convention used in experiment (Fig. 1).

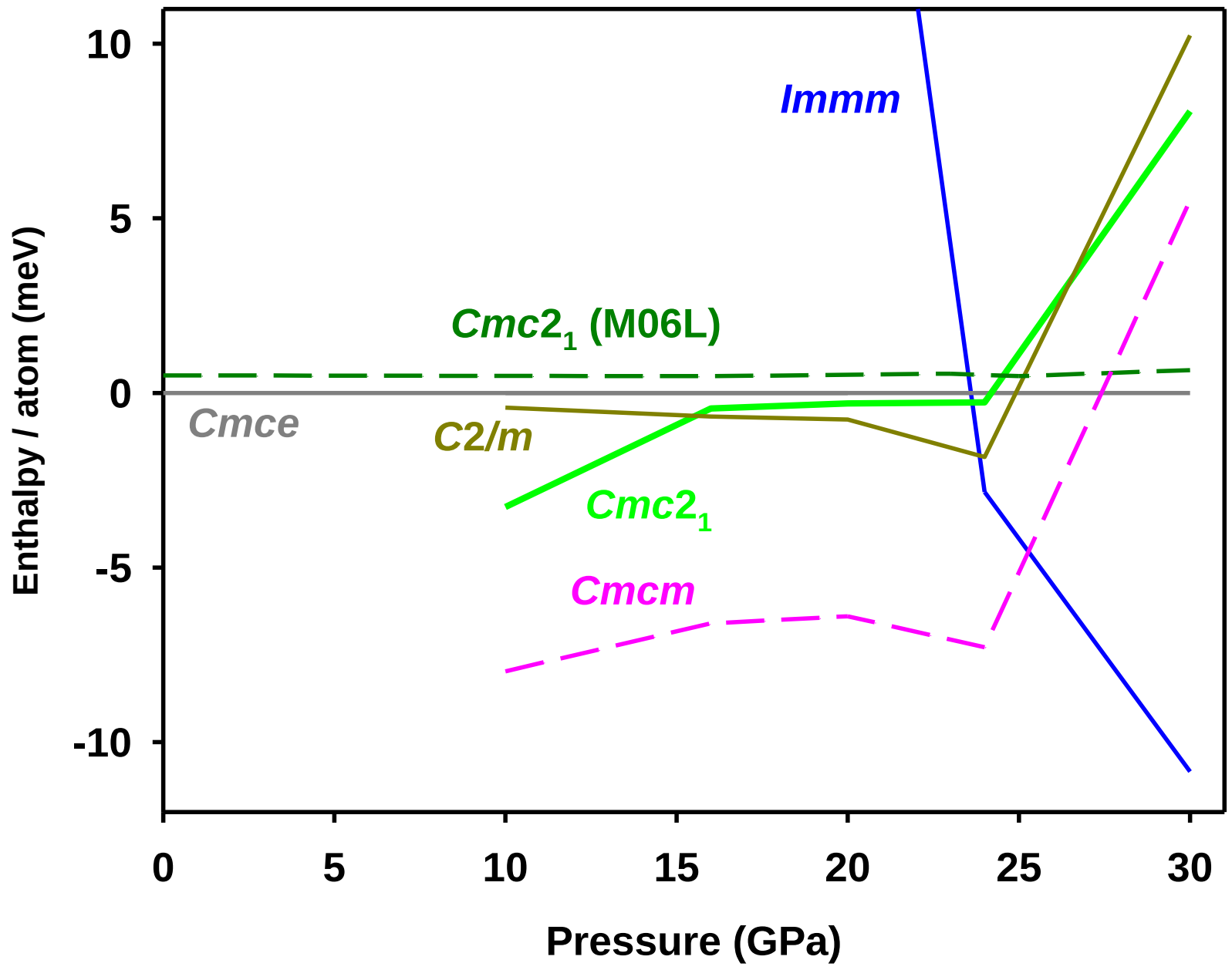

**Fig. S4.** Theoretically computed enthalpies of the candidate phases vs pressure determined here from the first principles plotted with respect to results for *Cmce* phase. Solid lines are the results computed using GGA-PBE functional, while a dashed line for *Cmc*$2_1$ phase is obtained using Minnesota 2006 local functional (M06-L). The computed equilibrium phases are dynamically stable in the presented here pressure range.

**Cmc2₁, 20 GPa**

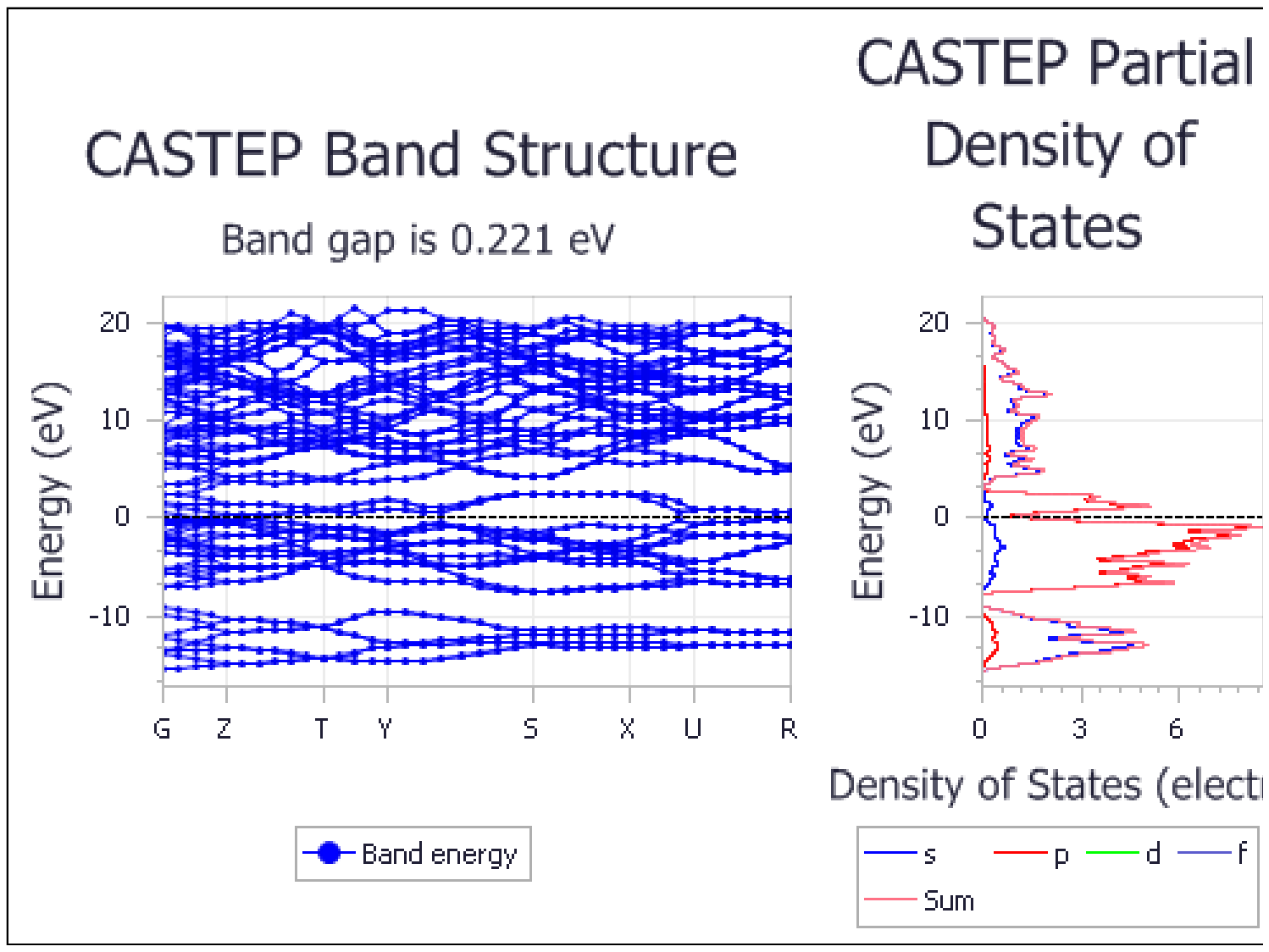

**Cmce, 20 GPa**

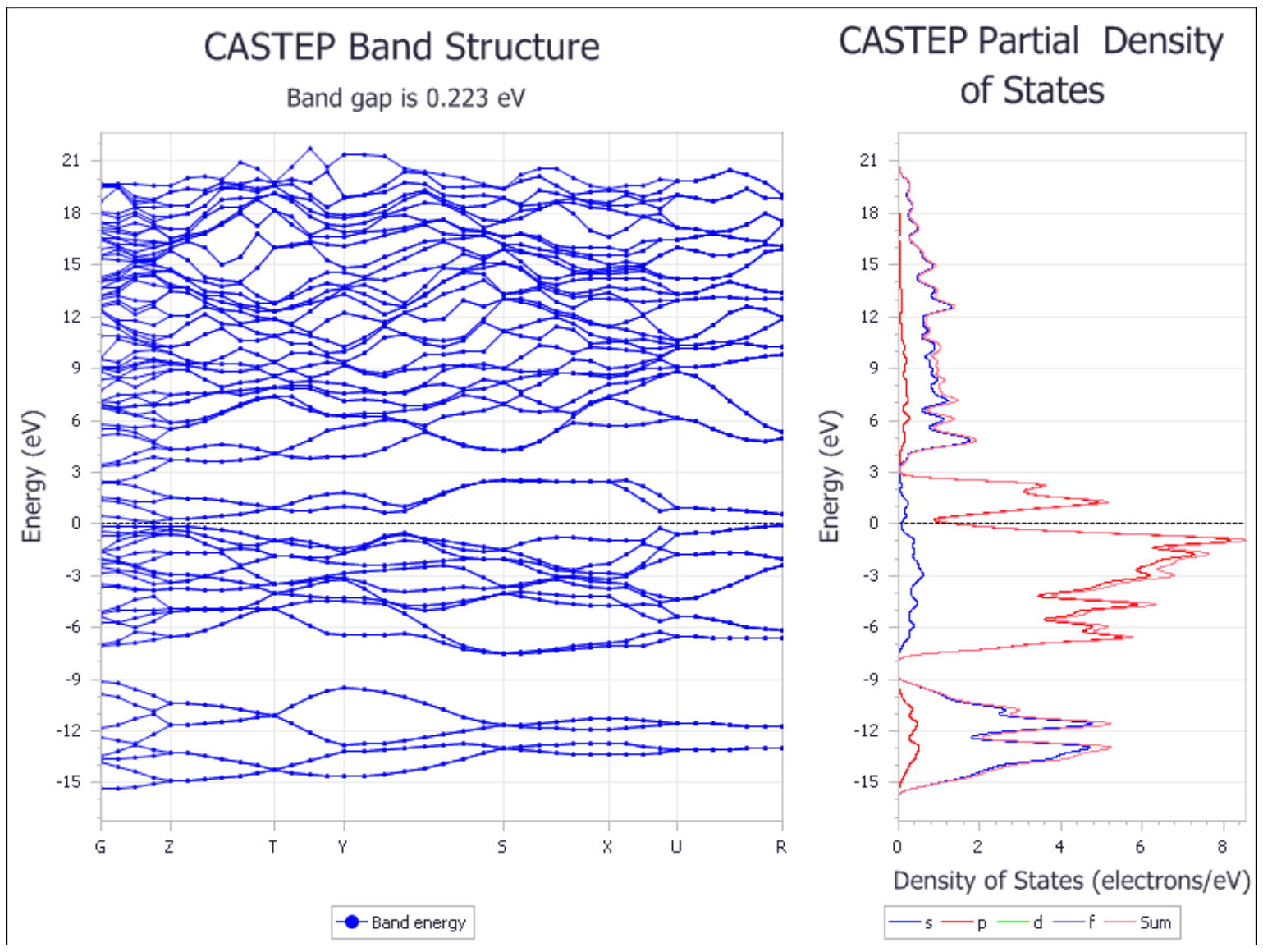

**Fig. S5.** Electronic band structure and density of electronic states of phases I (*Cmce*) and VI (*Cmc2₁*) at 20 GPa.

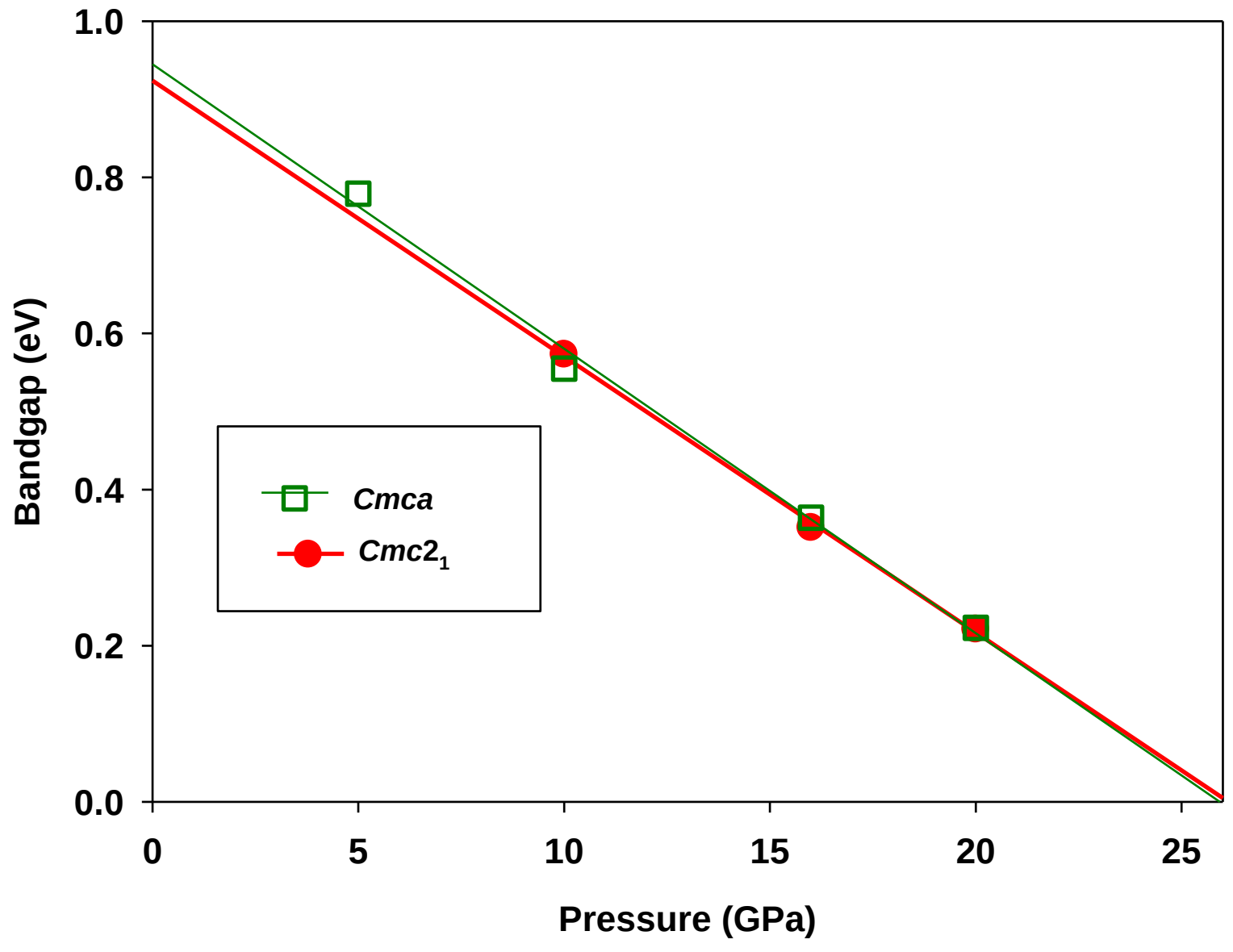

**Fig. S6.** The electronic band gap calculated as a function of pressure in *Cmce* and *Cmc*2$_1$ structures.

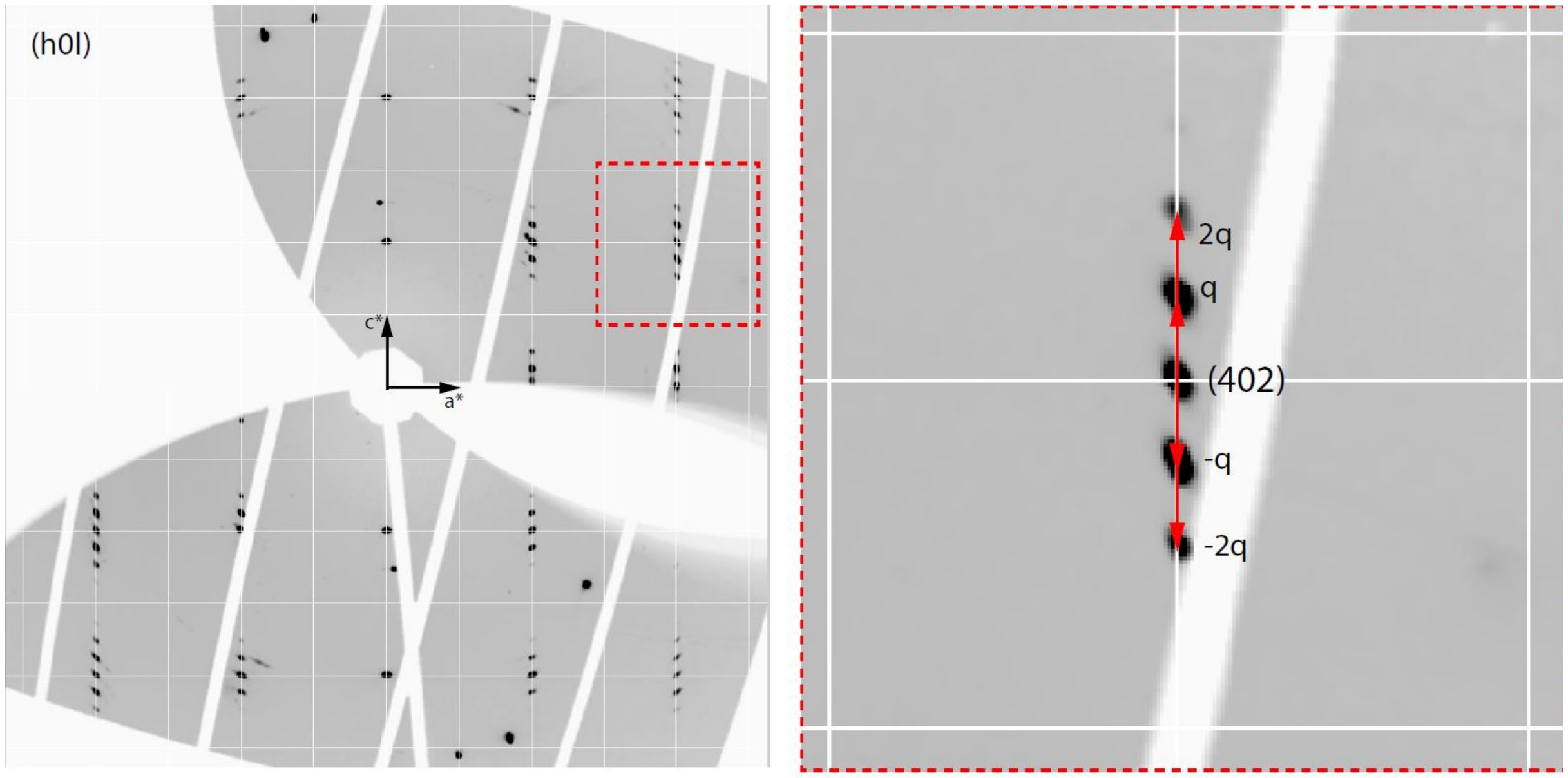

**Fig. S7.** Reconstructed reciprocal lattice planes of iodine phase V at 26.4 GPa. The right panel shows an enlarged view of a fine structure near the (402) reflection of a parent *Fmmm* phase. Satellite reflections are indexed in terms of the modulation vector. The details of the structural refinement are presented in Table S3 of Supplementary Materials).

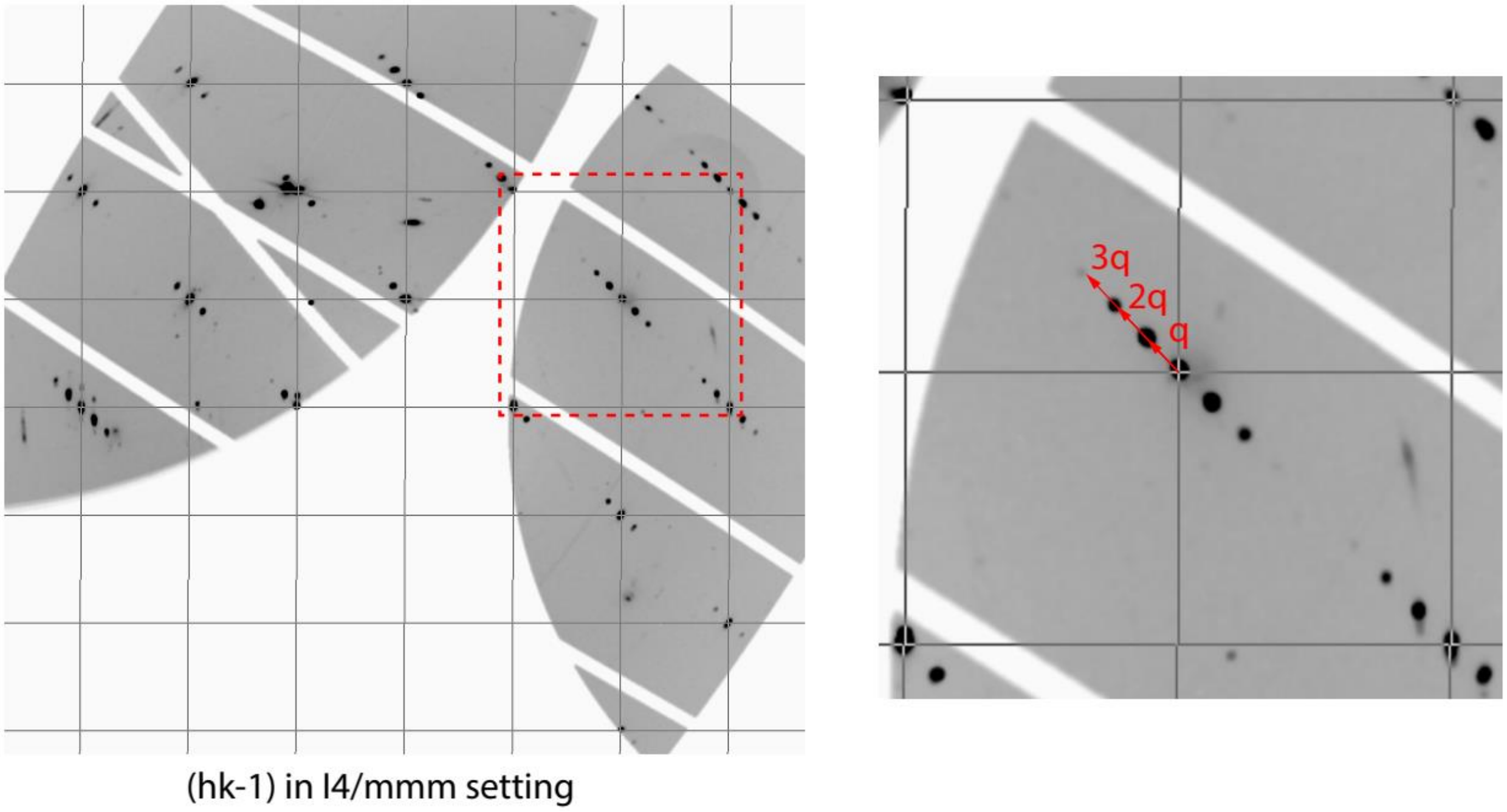

**Fig. S8.** Reconstructed precession image of iodine phase V at 26.4 GPa. The right panel shows an enlarged view of satellite reflections. Satellite reflections are indexed in terms of the modulation vector. The details of the structural refinement are presented in Table S3 of Supplementary Materials).

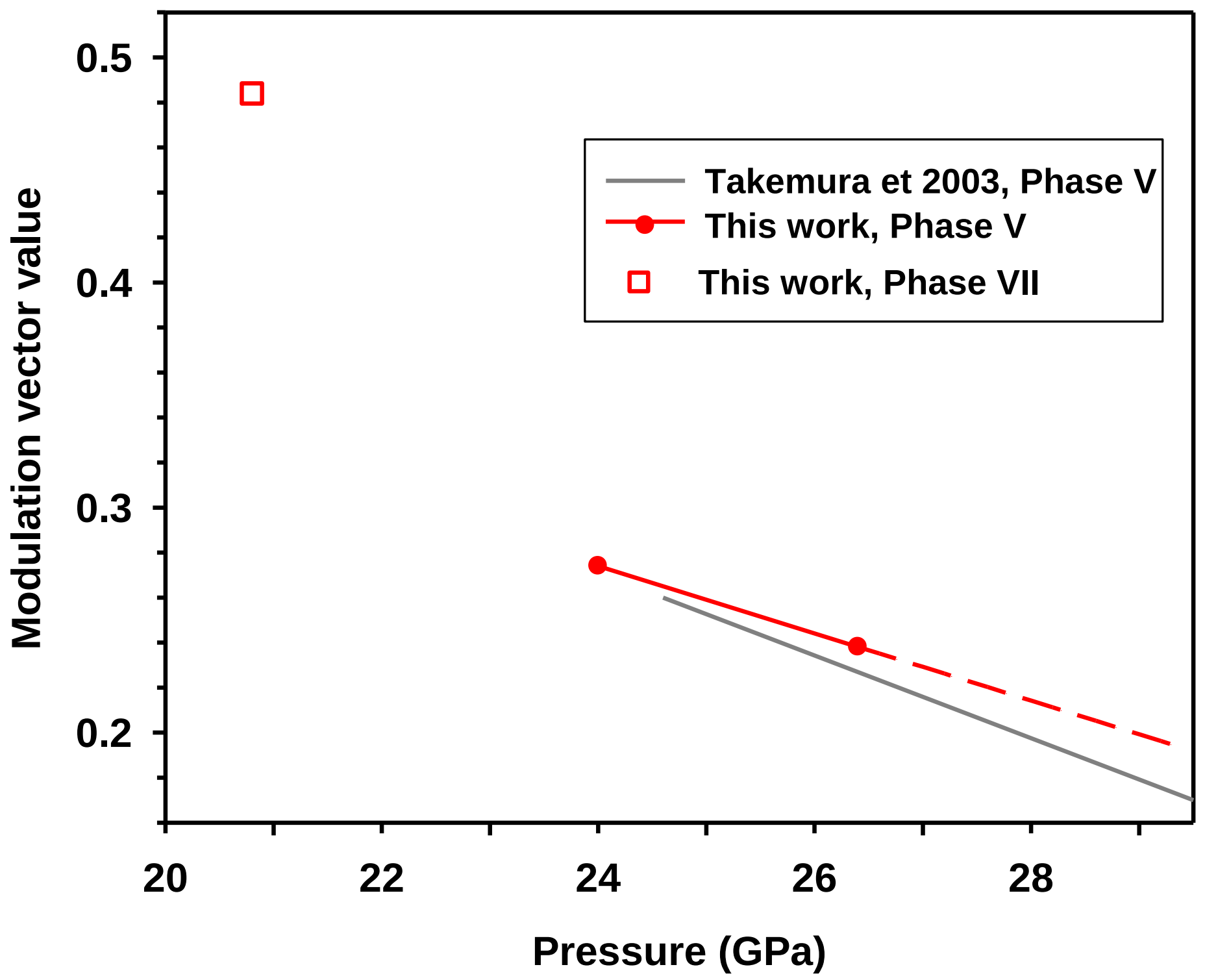

**Fig. S9.** The modulation vector value as a function of pressure determined here and Ref. [12].

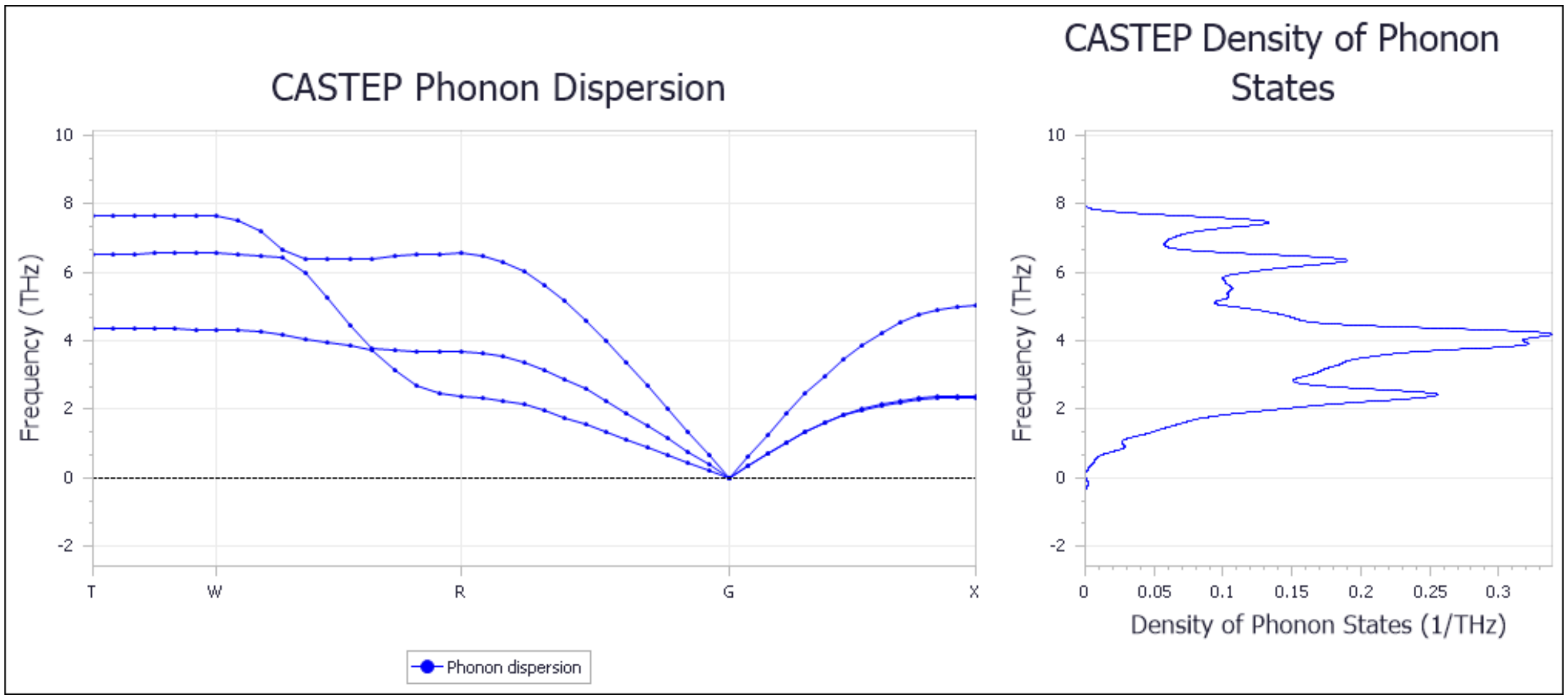

**Fig. S10.** Phonon dispersion curves and density of phonon states of phase II (*Immm*) at 30 GPa.

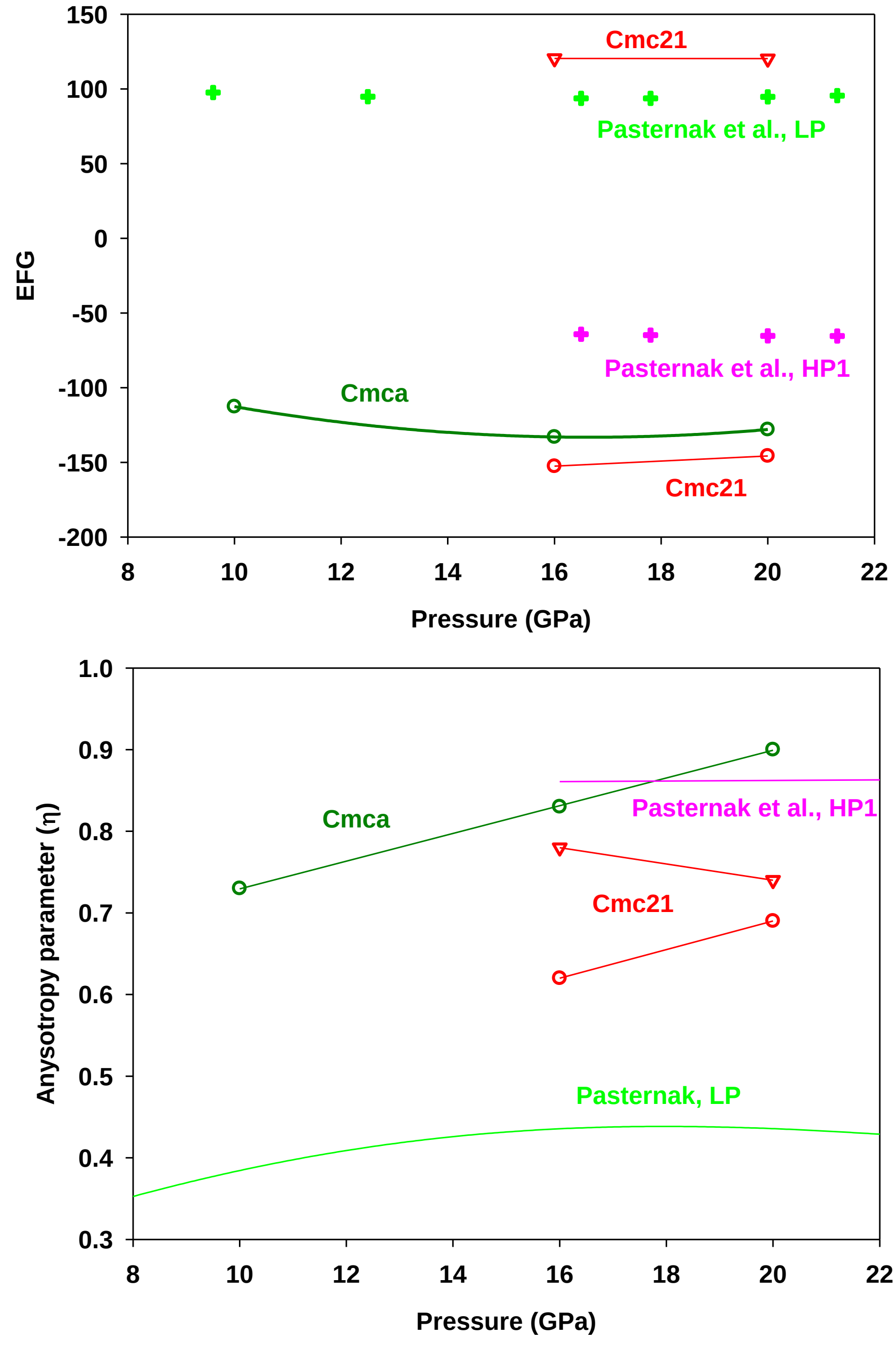

**Fig. S11.** Electric field gradient (EFG) parameters calculated for phases *Cmce* (phase I) and $Cmc2_1$ (phase VI) compared to experimental data of Ref. [16].

**Table S1. Crystal structure details of commensurate phases of iodine**

| Details of crystal structure refinements for iodine phases at high pressures* | | | | | |
|---|---|---|---|---|---|
| Pressure, GPa | 9.90 | 15.0 | 17.0 | 19.0 | 32.0 |
| Space group | *Cmce* | *Cmce* | $Cmc2_1$ | $Cmc2_1$ | *Immm* |
| Phase | iodine-I | iodine-I | iodine-VI | iodine-VI | iodine-II |
| Pressure step | p1 | p2 | p3 | p4 | p8 |
| *a* (Å) | 6.163(4) | 5.978(4) | 5.908(5) | 5.829(6) | 2.9077(6) |
| *b* (Å) | 4.1435(3) | 4.0396(4) | 4.0055(3) | 3.9773(6) | 3.0333(12) |
| *c* (Å) | 9.2864(8) | 9.1487(9) | 9.0992(11) | 9.0739(13) | 5.226(7) |
| *Z* | 2 | 2 | 2 | 2 | 2 |
| $V$ (Å$^3$) | 237.15(16) | 220.94(15) | 215.3(2) | 210.4(2) | 46.10(7) |
| $\rho_{calc}$ (g/cm$^3$) | 7.109 | 7.63 | 7.83 | 8.014 | 9.143 |
| μ/mm$^{-1}$ | 13.804 | 20.519 | 15.204 | 15.563 | 17.755 |
| $2\Theta_{min}$ for data collection (°) | 2.624 | 3.549 | 3.158 | 2.575 | 6.508 |
| $2\Theta_{max}$ for data collection (°) | 14.627 | 15.055 | 15.001 | 14.776 | 13.102 |
| Completeness to *d* = 0.8 Å | 0.527 | 0.468 | 0.432 | 0.545 | 0.25 |
| Reflections collected | 259 | 165 | 243 | 327 | 21 |
| Independent reflections | 136 | 86 | 175 | 213 | 16 |
| Independent reflections [$I > 2\sigma(I)$] | 132 | 80 | 171 | 213 | 15 |
| Refined parameters | 7 | 7 | 13 | 13 | 4 |
| $R_{int}(F^2)$ | 0.012 | 0.0131 | 0.0124 | 0.0087 | 0.083 |
| $R(\sigma)$ | 0.0148 | 0.0097 | 0.0163 | 0.0126 | 0.0316 |
| $R_1$ [$I > 2\sigma(I)$] | 0.0356 | 0.0246 | 0.031 | 0.0351 | 0.0836 |
| $wR_2$ [$I > 2\sigma(I)$] | 0.1028 | 0.0606 | 0.0784 | 0.0922 | 0.1911 |
| $R_1$ | 0.0368 | 0.026 | 0.0315 | 0.0351 | 0.0826 |
| $wR_2$ | 0.1057 | 0.0626 | 0.0785 | 0.0922 | 0.1907 |
| Goodness of fit on $F^2$ | 1.201 | 1.139 | 1.1 | 1.164 | 1.308 |
| $\Delta\rho_{max}$($e$ / Å$^3$) | 1.774 | 1.068 | 1.463 | 2.02 | 2.874 |
| $\Delta\rho_{min}$($e$ / Å$^3$) | -1.579 | -0.825 | -1.096 | -1.807 | -2.468 |
| *x*(I 1) | 0 | 0 | 0 | 0 | 0 |
| y(I 1) | 0.18376(9) | 0.19087(6) | 0.4299(2) | 0.4306(3) | 0 |
| *z*(I 1) | 0.62197(4) | 0.62278(3) | 0.26446(10) | 0.26390(15) | 0 |
| $U_{eq}$(I 1) (Å$^2$)* | 0.0165(6) | 0.0099(6) | 0.0080(9) | 0.0086(9) | 0.011(5) |
| *x*(I 2) | | | 0 | 0 | |
| y(I 2) | | | 0.0387(3) | 0.0291(6) | |
| *z*(I 2) | | | 0.51173(10) | 0.51229(15) | |
| $U_{eq}$(I 2) (Å$^2$)* | | | 0.0096(9) | 0.0114(10) | |
| *$U_{eq}$ is defined as one third of the trace of the orthogonalized $U_{ij}$ tensor. | | | | | |

**Table S2. Crystal structure details of incommensurate *Fmmm*(00γ)*s*00 phase VII of iodine.**

| | |
|---|---|
| **Crystal data** | |
| Chemical formula | I |
| $M_r$ | 126.9 |
| Crystal system, space group | Orthorhombic, *Fmmm*(00γ)*s*00† |
| Pressure step | P05 |
| Pressure (GPa) | 20.8 GPa |
| Wave vectors | **q** = 0.4837(2)**c*** |
| *a*, *b*, *c* (Å) | 3.9536(14), 5.735(18), 4.530(2) |
| *V* (Å$^3$) | 102.7(3) |
| *Z* | 4 |
| Radiation type | X-ray, λ = 0.2952 Å |
| μ (mm$^{-1}$) | 16.191 |
| **Data collection** | |
| Diffractometer | GSECARS 13IDD |
| No. of main reflections: All/independent/observed independent | 75/35/31 |
| No. of satellite reflections m = 1: All/independent/observed independent | 131/56/56 |
| No. of satellite reflections m = 2: All/independent/observed independent | 152/64/58 |
| $R_{int}$ | 0.0138 |
| $(\sin \theta/\lambda)_{max}$ (Å$^{-1}$) | 0.867 |
| **Refinement** | |
| $R(obs)_{main + sattelites}$ / $wR(all)_{main + satellites}$ | 0.0360/0.1141 |
| $R(obs)_{main}$ / $wR(all)_{main}$ | 0.0344/0.0934 |
| $R(obs)_{1st\ order\ satellites}$ / $wR(all)_{1st\ order\ satellites}$ | 0.0305/0.0812 |
| $R(obs)_{2nd\ order\ satellites}$ / $wR(all)_{2nd\ order\ satellites}$ | 0.0513/0.1538 |
| No. of reflections | 145 |
| No. of parameters | 6 |
| $\Delta\rho_{max}$, $\Delta\rho_{min}$ (e Å$^{-3}$) | 1.65, −1.13 |
| **Crystal structure** | |
| I (*x*,*y*,*z*) | (0, 0, 0) |
| $A_x^1$ | 0.0749(2) |
| $A_z^2$ | -0.00286(9) |

† Symmetry operations: (1) $x_1, x_2, x_3, x_4$; (2) $-x_1, -x_2, x_3, x_4+1/2$; (3) $-x_1, x_2, -x_3, -x_4$; (4) $x_1, -x_2, -x_3, -x_4+1/2$; (5) $-x_1, -x_2, -x_3, -x_4$; (6) $x_1, x_2, -x_3, -x_4+1/2$; (7) $x_1, -x_2, x_3, x_4$; (8) $-x_1, x_2, x_3, x_4+1/2$; (9) $x_1, x_2+1/2, x_3+1/2, x_4$; (10) $-x_1$,

$-x_2$+1/2, $x_3$+1/2, $x_4$+1/2; (11) $-x_1$, $x_2$+1/2, $-x_3$+1/2, $-x_4$; (12) $x_1$, $-x_2$+1/2, $-x_3$+1/2, $-x_4$+1/2; (13) $-x_1$, $-x_2$+1/2, $-x_3$+1/2, $-x_4$; (14) $x_1$, $x_2$+1/2, $-x_3$+1/2, $-x_4$+1/2; (15) $x_1$, $-x_2$+1/2, $x_3$+1/2, $x_4$; (16) $-x_1$, $x_2$+1/2, $x_3$+1/2, $x_4$+1/2; (17) $x_1$+1/2, $x_2$, $x_3$+1/2, $x_4$; (18) $-x_1$+1/2, $-x_2$, $x_3$+1/2, $x_4$+1/2; (19) $-x_1$+1/2, $x_2$, $-x_3$+1/2, $-x_4$; (20) $x_1$+1/2, $-x_2$, $-x_3$+1/2, $-x_4$+1/2; (21) $-x_1$+1/2, $-x_2$, $-x_3$+1/2, $-x_4$; (22) $x_1$+1/2, $x_2$, $-x_3$+1/2, $-x_4$+1/2; (23) $x_1$+1/2, $-x_2$, $x_3$+1/2, $x_4$; (24) $-x_1$+1/2, $x_2$, $x_3$+1/2, $x_4$+1/2; (25) $x_1$+1/2, $x_2$+1/2, $x_3$, $x_4$; (26) $-x_1$+1/2, $-x_2$+1/2, $x_3$, $x_4$+1/2; (27) $-x_1$+1/2, $x_2$+1/2, $-x_3$, $-x_4$; (28) $x_1$+1/2, $-x_2$+1/2, $-x_3$, $-x_4$+1/2; (29) $-x_1$+1/2, $-x_2$+1/2, $-x_3$, $-x_4$; (30) $x_1$+1/2, $x_2$+1/2, $-x_3$, $-x_4$+1/2; (31) $x_1$+1/2, $-x_2$+1/2, $x_3$, $x_4$; (32) $-x_1$+1/2, $x_2$+1/2, $x_3$, $x_4$+1/2.

Displacive modulations of I atom ($u_i(\overline{x_4})$ for $i$ = $x$, $y$, $z$) are described by a truncated Fourier series, which due to the symmetry restrictions take the following form: $u_x(\overline{x_4}) = A_x^1 \sin(2\pi\overline{x_4})$, $u_z(\overline{x_4}) = A_z^2 \sin(4\pi\overline{x_4})$

**Table S3. Crystal structure details of incommensurate *Fmmm*(00γ)*s*00 phase V of iodine.**

| **Crystal data** | | |
|---|---|---|
| Chemical formula | I | I |
| $M_r$ | 126.9 | 126.9 |
| Crystal system, space group | Orthorhombic, *Fmmm*(00γ)s00† | Orthorhombic, *Fmmm*(00γ)s00† |
| Pressure step | p6 | p7 |
| Pressure (GPa) | 24.0 | 26.4 |
| Wave vectors | **q** = 0.274(2)**c*** | **q** = 0.2381(6)**c*** |
| *a*, *b*, *c* (Å) | 4.2098 (13), 5.539 (10), 4.2407 (13) | 4.2018 (8), 5.434 (4), 4.2224 (8) |
| *V* (Å$^3$) | 98.88 (18) | 96.41 (8) |
| *Z* | 4 | |
| Radiation type | X-ray, λ = 0.2952 Å | |
| μ (mm$^{-1}$) | 16.82 | 17.25 |
| **Data collection** | | |
| Diffractometer | GSECARS 13IDD | GSECARS 13IDD |
| No. of main reflections All/independent/observed independent | 68/31/28 | 56/28/28 |
| No. of satellite reflections m = 1 All/independent/observed independent | 130/55/55 | 123/60/60 |
| No. of satellite reflections m = 2 All/independent/observed independent | - | 132/66/57 |
| $R_{int}$ | 0.017 | 0.012 |
| $(\sin \theta/\lambda)_{max}$ (Å$^{-1}$) | 0.842 | 0.876 |
| **Refinement** | | |
| $R_F(obs)_{\text{main + sattelites}}$ / $wR_F\ (all)_{\text{main + satellites}}$ | 0.0347/0.0419 | 0.0409/0.0515 |
| $R_F(obs)_{\text{main}}$ / $wRF\ (all)_{\text{main}}$ | 0.0257/0.0291 | 0.0400/0.0518 |
| $R_F(obs)_{\text{1st order satellites}}$ / $wRF\ (all)_{\text{1st order satellites}}$ | 0.0438/0.0499 | 0.0394/0.0486 |
| $R_F(obs)_{\text{2nd order satellites}}$ / $wRF\ (all)_{\text{2nd order satellites}}$ | - | 0.0468/0.0555 |
| No. of reflections | 86 | 154 |
| No. of parameters | 5 | 6 |
| $\Delta\rho_{max}$, $\Delta\rho_{min}$ (e Å$^{-3}$) | 1.84, −1.70 | 2.32, −3.05 |
| Crystal structure | | |
| I (*x*,*y*,*z*) | (0, 0, 0) | (0, 0, 0) |
| $A_x^1$ | 0.0576(4) | 0.0588(2) |
| $A_z^2$ | - | 0.00072(8) |

† Symmetry operations: (1) $x_1$, $x_2$, $x_3$, $x_4$; (2) $-x_1$, $-x_2$, $x_3$, $x_4$+1/2; (3) $-x_1$, $x_2$, $-x_3$, $-x_4$; (4) $x_1$, $-x_2$, $-x_3$, $-x_4$+1/2; (5) $-x_1$, $-x_2$, $-x_3$, $-x_4$; (6) $x_1$, $x_2$, $-x_3$, $-x_4$+1/2; (7) $x_1$, $-x_2$, $x_3$, $x_4$; (8) $-x_1$, $x_2$, $x_3$, $x_4$+1/2; (9) $x_1$, $x_2$+1/2, $x_3$+1/2, $x_4$; (10) $-x_1$, $-x_2$+1/2, $x_3$+1/2, $x_4$+1/2; (11) $-x_1$, $x_2$+1/2, $-x_3$+1/2, $-x_4$; (12) $x_1$, $-x_2$+1/2, $-x_3$+1/2, $-x_4$+1/2; (13) $-x_1$, $-x_2$+1/2, $-x_3$+1/2, $-x_4$; (14) $x_1$, $x_2$+1/2, $-x_3$+1/2, $-x_4$+1/2; (15) $x_1$, $-x_2$+1/2, $x_3$+1/2, $x_4$; (16) $-x_1$, $x_2$+1/2, $x_3$+1/2, $x_4$+1/2; (17) $x_1$+1/2, $x_2$, $x_3$+1/2, $x_4$; (18) $-x_1$+1/2, $-x_2$, $x_3$+1/2, $x_4$+1/2; (19) $-x_1$+1/2, $x_2$, $-x_3$+1/2, $-x_4$; (20) $x_1$+1/2, $-x_2$, $-x_3$+1/2, $-x_4$+1/2; (21) $-x_1$+1/2, $-x_2$, $-x_3$+1/2, $-x_4$; (22) $x_1$+1/2, $x_2$, $-x_3$+1/2, $-x_4$+1/2; (23) $x_1$+1/2, $-x_2$, $x_3$+1/2, $x_4$; (24) $-x_1$+1/2, $x_2$, $x_3$+1/2, $x_4$+1/2; (25) $x_1$+1/2, $x_2$+1/2, $x_3$, $x_4$; (26) $-x_1$+1/2, $-x_2$+1/2, $x_3$, $x_4$+1/2; (27) $-x_1$+1/2, $x_2$+1/2, $-x_3$, $-x_4$; (28) $x_1$+1/2, $-x_2$+1/2, $-x_3$, $-x_4$+1/2; (29) $-x_1$+1/2, $-x_2$+1/2, $-x_3$, $-x_4$; (30) $x_1$+1/2, $x_2$+1/2, $-x_3$, $-x_4$+1/2; (31) $x_1$+1/2, $-x_2$+1/2, $x_3$, $x_4$; (32) $-x_1$+1/2, $x_2$+1/2, $x_3$, $x_4$+1/2.

Displacive modulations of I atom ($u_i(\overline{x_4})$ for $i$ = $x$, $y$, $z$) are described by a truncated Fourier series, which due to the symmetry restrictions take the following form: $u_x(\overline{x_4}) = A_x^1 \sin(2\pi\overline{x_4})$, $u_z(\overline{x_4})$= $A_z^2 \sin(4\pi\overline{x_4})$

**Table S4. Vibrational modes and activity of *Cmce* and *Cmc*2$_1$ crystal structures**

| Space & point group | *Cmce* #64 ($D_{2h}$) | | *Cmc*2$_1$ #36 ($C_{2v}$) | | *Cmcm* #63 ($D_{2h}$) | |
|---|---|---|---|---|---|---|
| Site symmetry | 8f ($C_S$) | | 4a +4a ($C_S$) | | 4c ($C_{2v}$) + 4a ($C_{2h}$) | |
| Acoustic modes | $B_{1u}+B_{2u}+B_{3u}$ | | $A_1+B_1+B_2$ | | $B_{1u}+B_{2u}+B_{3u}$ | |
| Optical modes | Modes | Activity | Modes | Activity | Modes | Activity |
| | $A_u$ | Silent | $A_2$ | Raman | $A_u$ | Silent |
| | $B_{1u}+B_{2u}+B_{3u}$ | IR | $A_1+B_2+B_1$ | IR & Raman | $3B_{1u}+3B_{2u}+2B_{3u}$ (4$^a$) | IR |
| | $B_{1g}+B_{2g}$ | Raman (*a*) | $A_2+B_1$ | Raman & IR | none (4c) | |
| | $2A_g+2B_{3g}$ | Raman (*bc*) | $2A_1+2B_2$ | Raman & IR | $A_g+B_{1g}+B_{3g}$ | Raman |